\documentclass[A4,prl, superscriptaddress ,preprintnumbers,amsmath,amssymb,floatfix,footinbib,twocolumn]{revtex4}
\usepackage{graphicx}
\usepackage{epstopdf}
\usepackage{bm}
\usepackage{subfigure}
\usepackage{amsmath}
\usepackage{hyperref}

\newcommand{\D}{\mathcal{D}}

\newcommand{\bra}[1]{\langle#1\vert}
\newcommand{\ket}[1]{\vert#1\rangle}

\renewcommand{\vec}[1]{\mathbf{#1}}

\DeclareMathOperator{\Tr}{Tr}

\newcommand{\eqn}[1]{Eq.~\ref{#1}}
\newcommand{\fig}[1]{Fig.~\ref{#1}}

\renewcommand{\vec}[1]{\mathbf #1}

\newcommand{\footnoteremember}[2]{\footnote{#2}\newcounter{#1}\setcounter{#1}{\value{footnote}}}

\newcommand{\footnoterecall}[1]{\footnotemark[\value{#1}]}

\begin{document}

\title{
Dynamical Steady-States in Driven Quantum Systems 
}
\author{T.\ M.\ Stace}\email[]{stace@physics.uq.edu.au} 
\affiliation{ARC Centre for Engineered Quantum Systems, University of Queensland, Brisbane 4072, Australia}
\author{A.\ C.\ Doherty}
\author{D.\ J.\ Reilly}
\affiliation{ARC Centre of Excellence 
for Engineered Quantum Systems,
School of Physics, The University of Sydney, Sydney, NSW 2006, Australia}

\date{\today}

\begin{abstract}
We derive dynamical equations for a driven, dissipative quantum system in which the  environment-induced relaxation rate is comparable to the Rabi frequency, avoiding assumptions on the frequency dependence of the environmental coupling. 
When the environmental coupling varies significantly on the scale of the Rabi frequency,  secular or rotating wave approximations  
 break down.  Our approach avoids these approximations,  yielding  dynamical, periodic steady-states. This is important for the qualitative and quantitative description of the interaction between driven quantum dots and their phonon environment. 
The theory agrees well with recent experiments, 
 describing the transition from  asymmetric unsaturated resonances at weak driving to population inversion  at strong driving.

\end{abstract}

\maketitle

  The  basic physics of a dissipative, driven few-level system is understood via the Rabi or  Jaynes-Cummings models for a driven two-level atom \cite{PhysRevLett.107.100401} coupled to a dissipative bosonic environment, leading to the optical Bloch equations \cite{gar00}.  
  In the simplest case, the system response depends on the two-level  energy splitting, $\hbar\phi$, the  Rabi frequency, $\Omega$,  the environment-induced decay rate, $\Gamma$, and the driving frequency $\omega_0$.   
 When $\phi\gg\Omega,\Gamma$, various approximations lead to  \emph{Markovian} descriptions of the dynamics, in which correlations between the system and its external environment are very short lived.    Markovian models make distinct predictions depending on the relative size of $\Gamma$ and $\Omega$. If $\phi\gg\Gamma\gg\Omega$, 
the weak driving produces small Lorentzian resonances  in the system response, which are symmetric in the detuning $\tilde\eta=\phi-\omega_0$ \cite{barrett2006cmm,stace:136802,petta:186802}.  If $\phi\gg\Omega\gg\Gamma$, the strong driving  saturates the resonance, and excitation channels that are available  when $\omega_0>\phi$ lead to asymmetric resonances \cite{Dykman79,bra99b,stace:106801,Roulleau2011aa}.  


Driven, few-level double-quantum-dot (DQD) devices are promising building-blocks for a variety of new quantum technologies \cite{RevModPhys.79.1217}.  
In these devices, for which the electron-phonon coupling is significant \cite{hay03,Oosterkamp98Microwave_spectroscopy_of_a_quantum-dot_molecule,fuj98,petta:186802,Granger2012aa
}, $\Gamma$ or $\Omega$ may become large enough that Markovian approximations fail \cite{PhysRevLett.104.036801,JPSJ.74.3131,McCutcheon2010aa}. In this Letter, we derive a non-Markovian master equation that is valid in these circumstances. We first describe our general formalism, which we then exemplify with a  DQD coupled to a phonon bath.  We compute the response of the DQD to driving, and discuss the qualitatively new phenomena that emerge. 

  We assume the Hamiltonian {$H=H_S+H_I+H_B$}  for the driven system, interaction and harmonic  bath, where $H_I=Z_S (\sum_{\vec{q}} g_\vec{q} ^* 
   a_\vec{q}^\dagger+h.c)$,  $Z_S$ is a system operator,  $a_\vec{q}^\dagger$ is the bosonic bath creation operator for mode $\vec{q}$ and $g_\vec{q}$ is the system-bath coupling strength \cite{gar00,stace:106801}. Since the system is driven, $H_S$ has periodic time-dependence.  We transform to an interaction picture 
  with respect to $H_0=H_D+H_B$, where $H_D$ acts trivially on the bath, and has an associated set of Floquet eigenfrequencies $\mathcal{W}$ \cite{mar82}, giving $H(t)=H_{S}(t)+H_I(t)$, where  $ H_{S}(t)=\sum_{\omega\in\mathcal{W} }h_\omega e^{i \omega t}$,
\begin{equation}
H_I(t)= Z_S(t) \sum_{\vec{q}} g_{\vec{q}} ^* 
   a_{\vec{q}}^\dagger e^{i\omega_{\vec{q}}t}+h.c,\label{eqn:intham}
\end{equation}
 $  Z_S(t)= \sum_{\omega\in\mathcal{W} }P_{\omega} e^{i \omega t}$,  
%
  and the Fourier-coefficients $h_\omega$, $P_{\omega}$ satisfy $h_{-\omega}=h_{\omega}^\dagger$,  $P_{-\omega}=P_{\omega}^\dagger$.  It is common  to choose the dressing Hamiltonian $H_D=H_S$ so that $H_{S}(t)=0$. We avoid this,  to remove bath-induced dispersive shifts. 
  
To derive the system dynamics, we iterate the von Neumann equation for the joint system-bath density matrix, $R$,  then trace over the bath \cite{gar00}
\begin{equation}
\dot\rho(t)=-i[H_{S}(t),\rho(t)]-\Tr_B\!\!\int_{0}^t dt'  [H_I(t),[H_I(t'),R(t')]],\label{eq:vonNeumannmod}\nonumber
\end{equation}
where $\rho=\Tr_B R$ is the reduced system density matrix.
Various  approximations are often invoked to turn the integral  into a convolution with a rapidly decaying kernel, 
yielding Markovian evolution \cite{gar00}. Instead we take a Laplace transform which we solve using an ansatz  containing  the relevant poles of the problem.   
Truncating the set of poles yields tractable approximations.
 The Laplace transform is (see Supplementary Information [SI])
\begin{multline}
s \bar \rho_s-\rho(0)=-i\sum_{\omega'\in\mathcal{W}}[h_{\omega'},\bar\rho_{s-i\omega'}]\\
+ \sum_{\omega',\omega''\in\mathcal{W}}
i(\tilde J(\omega'+i s)-\tilde J(\omega''-i s))\,P_{\omega'}\bar\rho_{s-i(\omega'-\omega'')} P_{\omega''}^\dagger \\
{}-i \tilde J(\omega'+i s)\bar \rho_{s-i(\omega'-\omega'')}
P_{\omega''}^\dagger P_{\omega'} \\
+i \tilde J(\omega''-i s) 
P_{\omega''}^\dagger P_{\omega'}\bar\rho_{s-i(\omega'-\omega'')},\label{eqn:lt3}
\end{multline}
where $\bar \rho_s=\int_0^\infty dt' e^{-s t'}\rho(t')$, $\tilde J(x)=\sum_q \frac{|g_q|^2}{\omega_q+x}$ is the generalised {spectral density} and we assume that $R(t)\approx\rho(t)\otimes\rho_B$ for a thermal bath state, $\rho_B$, at zero temperature \footnote{It is straightforward to include finite temperatures in the formalism, see Supplementary Information.}.  For later reference we define $\hat J, J$ and $F$: 
\begin{equation}
\lim_{s\rightarrow 0^+}\pm i\tilde J(\omega\pm i s)=\hat J_{\pm}(\omega)=(J(\omega)\pm i F(\omega))/2,\label{eqn:spec}
\end{equation}
where $J(x)=2\pi \sum_\vec{q} |g_\vec{q}|^2\delta(\omega_\vec{q}+x)$ is the  spectral density  and 
 $F(x)=-\pi^{-1}\int d\omega \,J(\omega)/(\omega-x)$ \cite{gar00}\footnoteremember{spectdens}{This form of the spectral density reverses the common convention for which $J(\omega)=0$ for $\omega<0$.}[SI].

Since $\rho(t)$ is bounded,  Mittag-Leffler's theorem implies that $\bar\rho_s$ is  determined by its poles \cite{jeffreys1999methods}. 
Suppose $\bar \rho_s$ has a pole at $s=z$; consistency between the RHS and LHS of \eqn{eqn:lt3} then requires additional poles in $\bar \rho_s$ at $s=z+i(\omega'-\omega'')$, where $\omega',\omega''\in\mathcal{W}$.  This motivates the ansatz
\begin{equation}
\bar\rho_s=\sum_{\nu\in\mathcal{V}}\frac{\rho_\nu}{s-i\nu},
\label{eqn:nst}
\end{equation}
where $\rho_\nu$ are  as-yet-unknown residues.  For  consistency in \eqn{eqn:lt3}, $\mathcal{V}$ is a countably infinite set, with   $\mathcal{W}\subset\mathcal{V}$.    To make progress, we  truncate $\mathcal{V}$ to a finite  set of the most significant poles,
 and require that the residues of poles that appear on the LHS of \eqn{eqn:lt3} equal those on the RHS.  This becomes exact in the limit that a complete set of poles is retained. 

Since $\Tr \rho =1$, $\rho$ has a non-zero steady-state, 
  so $\bar\rho_s$ has a pole at $s=0$.  This suggests the simple but illuminating case in which we retain only this pole, (i.e.\ $\mathcal{V}=\{0\}$) so $\bar\rho_s=\rho_0/s$, and the LHS of \eqn{eqn:lt3} becomes a constant (i.e.\ its residue is 0).  The residue of the RHS at $s=0$ should therefore vanish, yielding 
\begin{equation}
0=-i[h_{0}-f_0,\rho_0]+\sum_{\omega\in\mathcal{W}} J(\omega)\D[P_\omega]\rho_{0},\label{eqn:RWA}
\end{equation}
where $f_0=\sum_{\omega}F(\omega) P_{\omega}^\dagger P_{\omega}/2$ and $\D[A]\rho\equiv A\rho A^\dagger-(A^\dagger A \rho+\rho A^\dagger A)/2$ is the Lindblad superoperator. 
The solution to \eqn{eqn:RWA} is the 
 steady-state of the conventional  Markovian dynamics \cite{gar00,stace:106801}.  Furthermore, $F$ yields a dispersive Lamb shift that renormalises the system dynamics.  
Choosing $H_D$ such that $h_0=f_0$  cancels the dispersive effects arising from the bath, so that the \emph{renormalised} system Hamiltonian vanishes in this interaction picture. 

In general,  consistency between  residues appearing on  the LHS and on the RHS of \eqn{eqn:lt3} as $s\rightarrow i\nu'$ requires 
\begin{multline}
i\nu'\rho_{\nu'}=
-i\sum_{\nu\in\mathcal{V}}[h_{\nu'-\nu}, \rho_{\nu}]\\
{}+
\sum_{
\tiny\begin{matrix}\nu\!\in\!\mathcal{V},\omega\!\in\!\mathcal{W}, \\ {\omega'\!=\!\omega\!+\!\nu\!-\!\nu'}\end{matrix} 
}\hspace{-0.2cm}\Big(
(\hat J_+(\omega-\nu')+ \hat J_-(\omega+\nu))\,P_{\omega}\rho_{\nu} P_{\omega'}^\dagger \\
{} -( \hat J_{\tiny{+}}(\omega-\nu') \rho_{\nu}
P_{\omega'}^\dagger P_{\omega}
+ \hat J_-(\omega+\nu) 
P_{\omega'}^\dagger P_{\omega}\rho_{\nu})\Big).\label{eqn:residues2}
\end{multline}
The residues appearing in \eqn{eqn:nst} are bounded matrices, so we see from \eqn{eqn:residues2} that $\|\rho_{\nu'}\|\sim |\hat J/{\nu'}|$, 
i.e.\ the size of the residue decreases with the magnitude of the pole. 
As above, we choose  $H_D$  so that $h_\omega=f_\omega$ to cancel dispersive  terms in \eqn{eqn:residues2}.  This choice fixes the poles that appear in $\mathcal{W}$. 
The  $\hat J$-dependent terms arising on the RHS of \eqn{eqn:residues2} can be written as a sum of (a) dissipative $J$-dependent terms, analogous to the Lindblad terms in \eqn{eqn:RWA}, (b) dispersive terms of the form $i[f_{\nu'-\nu}, \rho_{\nu}]$ where $f_\nu$ depends on $F$, and are eliminated by the correct choice of $H_D$ analogous to $f_0$ in \eqn{eqn:RWA}, 
and (c) residual inhomogenous $F$-dependent terms, which cannot be eliminated.

To generate   transient dynamics of a system we should retain poles with negative real values. 
  In what follows, we will  be concerned with steady-state properties of a system operator $M(t)=\sum_{\omega\in\mathcal{W}}M_\omega e^{i\omega t}$, and so we  consider only pure-imaginary poles.  The Laplace transform is $
\bar M_s=\sum_{\omega\in\mathcal{W}}\Tr\{M_\omega\bar\rho_{s-i\omega}\}
$ and the time-averaged, steady-state expectation is the residue of $\bar M_s$ at $s=0$, i.e.\ $\langle M\rangle_0\equiv\overline{\Tr\{M(t)\rho(t)\}}=\sum_{\nu\in\mathcal{V}} \Tr\{M_\nu^\dagger\rho_\nu\}$. 
Importantly, this depends on both the time-averaged steady state $\rho_0$, and the dynamical residues $\rho_{\nu\neq0}$.  

We note  that the dynamical poles in the ansatz yields non-Markovian evolution: in the dressed basis, Markovian dynamics leads to stationary steady-states, which implies $\rho_{\nu\neq0}=0$.  This corresponds to the simplest approximation, $\mathcal{V}=\{0\}$, discussed above.    

    

Using this  formalism, we now turn to the example of a microwave-driven, one-electron DQD system with localised left/right states $\ket{l},\ket{r}$, separated by a distance $d$, inter-dot bias, $\epsilon$, and inter-dot tunnelling rate $\Delta$, driven at frequency $\omega_0$ and amplitude $\Omega_0$, coupled to a phonon bath \cite{barrett2006cmm,stace:106801,stace:205342}.   The driven system Hamiltonian is 
\begin{subequations}
\begin{align}
H_S&=-(\epsilon \sigma_z+\Delta\sigma_x )/{2} + {\Omega_0} \cos( \omega_0 t) 
  (\cos\delta\,\sigma_z +\sin \delta \,\sigma_x ),\nonumber\\
  &=-\phi\sigma_z^e/2+\Omega_0\cos(\omega_0 t)(\cos(\theta-\delta)\sigma_z^e -\sin(\theta-\delta)\sigma_x^e),\nonumber
\end{align}
\end{subequations}
where $\sigma_z\equiv\ket{l}\bra{l}-\ket{r}\bra{r}$ and $\sigma_x\equiv\ket{l}\bra{r}+\ket{r}\bra{l}$.  The qubit splitting is $\phi=(\epsilon^2+\Delta^2)^{1/2}$ in the energy eigenbasis $\{\ket{g},\ket{e}\}$, $\sigma_z^{e}=\sin \theta \,\sigma_x + \cos \theta \,\sigma_z$, $\sigma_x^{e}=\cos \theta \,\sigma_x - \sin \theta \,\sigma_z$ and $\theta = \arctan(\Delta/\epsilon)$.
In a frame rotating at the driving frequency $\omega_0$ (and dropping terms with frequency $\pm2\omega_0$ \footnote{This is not critical to the approach and can be corrected by including more Floquet terms \cite{mar82,PhysRevLett.107.100401}.}) $H_S$ becomes  time-independent\begin{equation}
 H_S=-(\tilde \eta \sigma_z^e+\tilde \Omega \sigma_x^e)/2\equiv-\tilde\Omega'(\cos\tilde\varphi \,\sigma_z^e+\sin\tilde\varphi \,\sigma_x^e)/2,\nonumber
\end{equation}
where $\tilde\eta=\phi-\omega_0$, $\tilde\Omega=\Omega_0\sin(\theta-\delta)$, $\tilde\varphi=\arctan(\tilde\Omega/\tilde\eta)$ and $\tilde \Omega'=({\tilde\Omega^2+\tilde\eta^2})^{1/2}$  (here $\tilde{}$ denotes bare quantities).  

We transform to an interaction picture defined by $$H_D=-( \eta \sigma_z^e+ \Omega \sigma_x^e)/2=- \Omega' \sigma_z^d/2,$$
in the dressed basis $\{\ket{-},\ket{+}\}$,  $\sigma_z^{d}=\cos\varphi \,\sigma_z^e+\sin\varphi \,\sigma_x^e$,
 $\varphi=\arctan(\Omega/\eta)$ and $ \Omega'=({\Omega^2+\eta^2})^{1/2}$.    This gives the Fourier coefficients  of  $H_S(t)$ defined earlier;  $h_0=-(\tilde \Omega'\cos(\tilde \varphi-\varphi)-\Omega')\sigma_z^d/2$ and 
 $h_{\pm\Omega'}=-\tilde\Omega'\sin(\tilde \varphi-\varphi)\sigma_\pm^d/2$.  We emphasise that  $\eta$ and $\Omega$ in $H_D$ 
 will  be chosen below to cancel bath-induced dispersive terms.
 
The electron-phonon coupling 
Hamiltonian is given by \eqn{eqn:intham} with $Z_S=\sigma_z$ \cite{stace:106801,mahan,rahman2007pulse}.   $Z_S(t)$ has Fourier frequencies $\mathcal{W}=\pm\{0, \Omega', \omega_0,  \omega_0\pm\Omega'\}$ and coefficients 
 $P_0=\alpha_0 \sigma^{d}_z $, 
$  P_{\Omega'} = \alpha_{\Omega'}\sigma^{d}_+$, \mbox{$
  P_{\omega_0 \pm\Omega'} =\alpha_{\omega_0 \pm\Omega'}  \sigma^{d}_\pm$, 
  $P_{\omega_0}= \alpha_{\omega_0} \sigma^{d}_z$}, where  
  $\alpha_0=\cos \theta \cos \varphi$, $\alpha_{\Omega'}=- \cos \theta \sin \varphi$, $\alpha_{\omega_0 \pm\Omega'} =\mp \sin \theta \left(1\pm\cos \varphi \right)/2$, $\alpha_{\omega_0}= -\sin \theta \sin \varphi/2$    \cite{stace:106801}.

\begin{figure}[t]
\begin{center}
\includegraphics[width=\columnwidth]{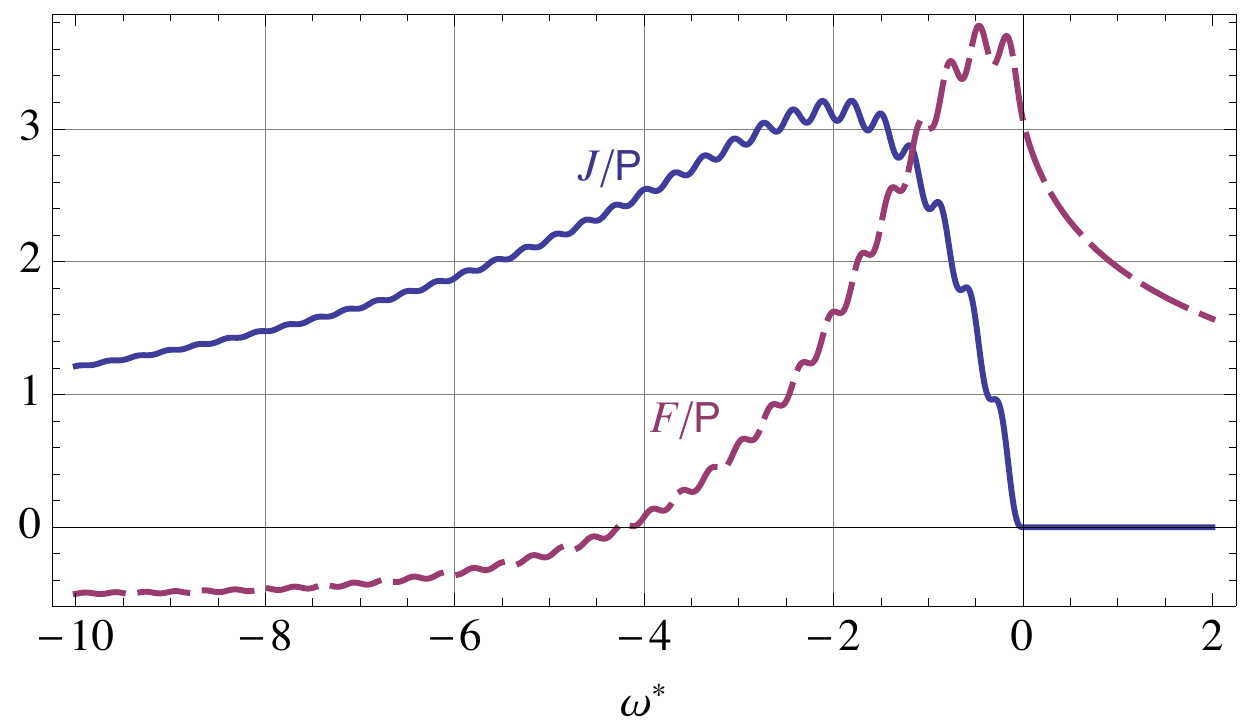}
\caption{Dependence of $J$ (solid) and $F$ (dashed) on $\omega$, for $\omega_c^*=2$ and $d^*=20$.  
} \label{fig:JF}
\end{center}
\end{figure}
    
To illustrate the significance of the dynamical poles, we choose $\mathcal{V}=\{0,\pm\Omega'\}$. Note that $\Omega'$ is the \emph{renormalised} Rabi frequency, which is fixed once $H_D$ is chosen to eliminate dispersive shifts. 
The dispersive terms appearing in \eqn{eqn:residues2} are $f_0=a_0\,\sigma_z^d/2$,  and $f_{\pm\Omega'}=a_{\Omega'}\,\sigma_{\pm}^d/2$  where
%
\begin{subequations}
\begin{align}
a_0&=(-\alpha_{\omega_0-\Omega'}^2 F_{\omega_0-\Omega'}+\alpha_{\omega_0+\Omega'}^2 F_{\omega_0+\Omega'}+\alpha_{\Omega'}^2 F_{\Omega'})/2,\nonumber\\
a_{\Omega'}&=\alpha _{\omega _0} (\alpha _{\omega _0-\Omega' } F_{\omega _0-\Omega' }-\alpha _{\omega _0+\Omega' } F_{\omega
   _0+\Omega' })-
   \alpha _0 \alpha _{\Omega' }
   F_{\Omega' },\nonumber
   \end{align}
\end{subequations}
and $F_x\equiv F(x)-F(-x)$.  
Setting $h_\nu=f_\nu$ to cancel dispersive terms yields 
  the required relationship between the bare $\tilde\eta,\tilde\Omega$ and the renormalised $\eta,\Omega$,
\begin{equation}
\left[\begin{array}{c} \tilde\eta \\ \tilde \Omega\end{array}\right]=-\left[\begin{array}{cc}\cos\varphi & -\sin\varphi \\\sin\varphi & \cos\varphi\end{array}\right].
\left[\begin{array}{c} a_0-\Omega' \\ a_{\Omega'}\end{array}\right].\label{eqn:nonlin}
\end{equation}
We solve this nonlinear equation numerically for $\eta,\Omega$. 

The renormalisation of the detuning arises from the phonon-induced Lamb shift \cite{gar00}.  This shift  depends on the phonon modes with which the driven system is most strongly coupled, which depends itself on  detuning \cite{stace:106801}, resulting in a detuning-dependent Lamb shift.  The renormalisation of the Rabi frequency is related to the polaron transformation \cite{bra99b,PhysRevLett.110.217401}, in which pure-electronic modes are renormalised to polaronic modes with larger effective mass, reducing the transition dipole moments.

 Differences between the bare and renormalised quantities ultimately depend on the electron-phonon coupling via $F$, which appears in $a_0$ and $a_{\Omega'}$.  
Very close to resonance, $\tilde \eta\approx0$, some intuition into the effect of the dispersive shifts can be gained by expanding the RHS of \eqn{eqn:nonlin} in powers of $\Omega'$ then solving to find 
\begin{subequations}\label{eqn:approxetaomega}
\begin{align}
\eta_{\textrm{approx}}&\approx{\tilde \eta}/({1+ F'(0)})+O[{\tilde \eta^3}/{\tilde \Omega^2}]
,\label{eqn:approxetaomegaA}\\
 \Omega_{\textrm{approx}}&\approx {\tilde \Omega}/({1- F'(0)})
 +O[{\tilde \eta^2}/{\tilde \Omega}].
 \label{eqn:approxetaomegaB}
 \end{align}
\end{subequations}
Thus, the renormalised detuning and Rabi frequency are scaled with respect to the bare values.  Importantly, the scaling factors depend (non-perturbatively) on $F'(0)$.

\begin{figure}
\begin{center}
\includegraphics[width=\columnwidth]{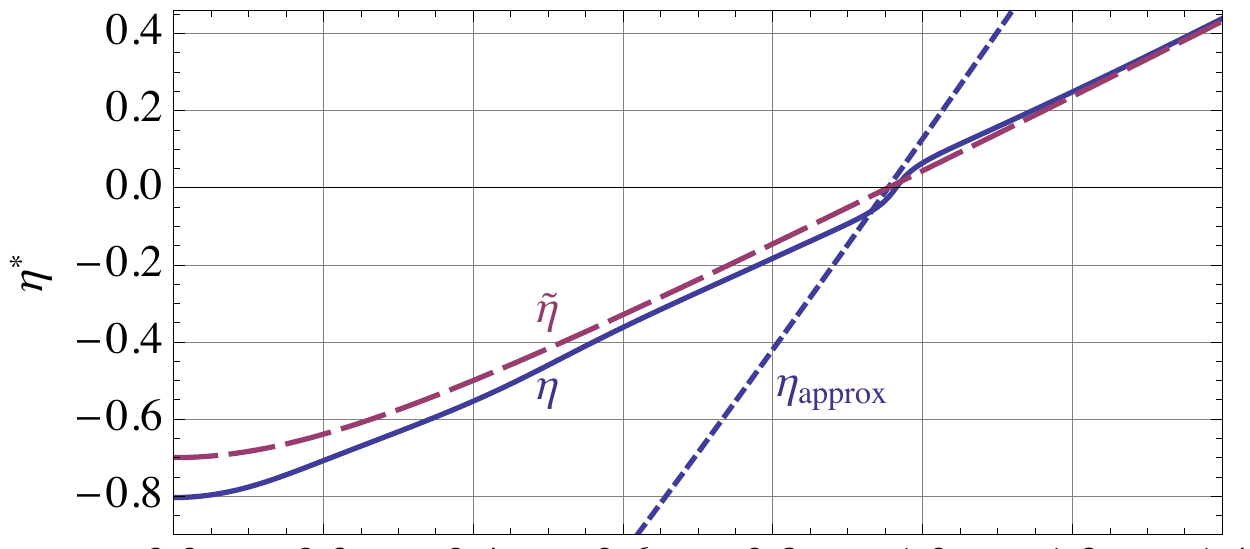}
\includegraphics[width=\columnwidth]{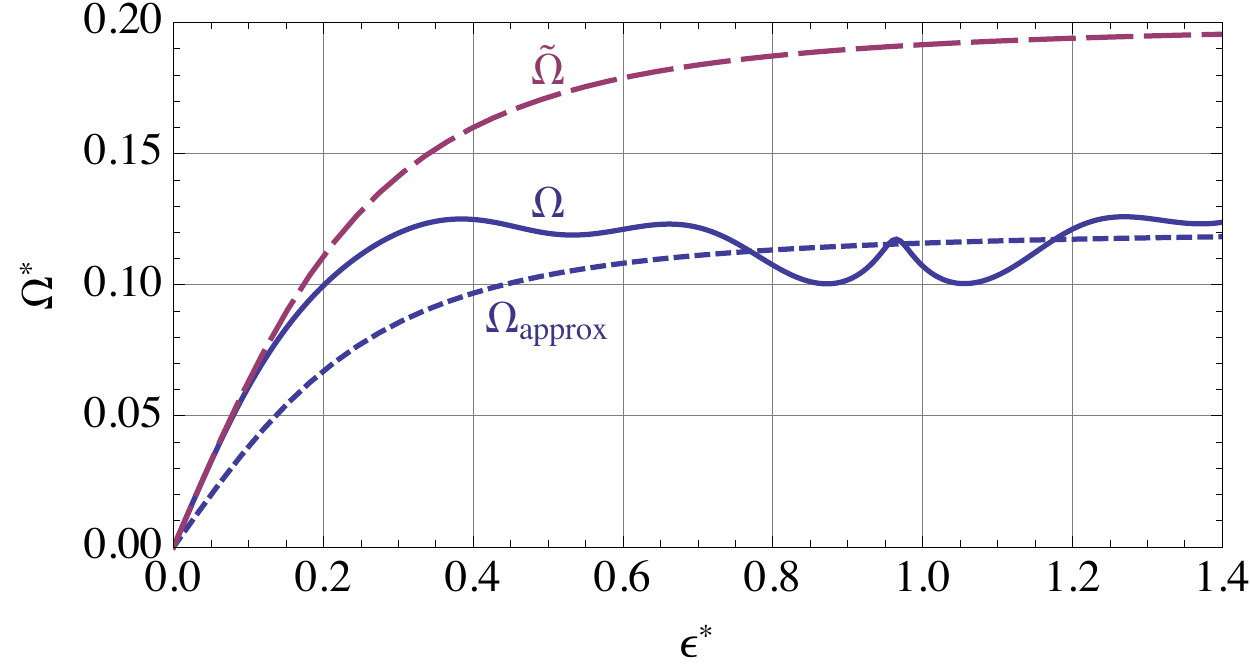}
\caption{Top: Detuning versus bias. Bottom: Rabi frequency versus bias.   Dashed curves: Bare  detuning $\tilde \eta$ and  Rabi frequency $\tilde \Omega$. Solid curves: Renormalised detuning $\eta$ and Rabi frequency $\Omega$ obtained by solving \eqn{eqn:nonlin}. Dotted curves: Near resonance approximations to renormalised quantities from \eqn{eqn:approxetaomega}; $\eta_{\textrm{approx}}$ captures the steeper slope of $\eta$ close to resonance.  Parameters are $\Delta^*=0.3,  \delta=\pi/2, \omega_c^*=2, d^*=20, \Omega_0^*=0.2, \mathsf{P}=0.2$.  Resonance occurs at $\epsilon^*=(1-{\Delta^*}^2)^{1/2}\approx0.954$.
} \label{fig:etaomega}
\end{center}
\end{figure}

Focussing on  the specific case of bulk piezo-electric phonon coupling, the  spectral density is \cite{stace:106801} 
\begin{equation}
J(\omega^*)=\begin{cases}
    \pi\mathsf{P}|\omega^*|\frac{1-\mathrm{sinc}(d^* \omega^*)}{1+(\omega^*/\omega_c^*)^2} & \text{if } \omega^* < 0 \\
   0       &\text{if }  \omega^* \geq 0
  \end{cases}
 .\label{eqn:J}
\end{equation}
where $\omega^*\equiv\omega/\omega_0,  d^*\equiv d \,\omega_0/c_s$ and $\mathsf{P}$ are, respectively, nondimensionalised frequency, inter-dot separation, and coupling strength ($c_s$ is the speed of sound) \footnoterecall{spectdens}. The high-energy cutoff, $\omega_c^*$, is determined by the spatial extent of the localised wavefunctions $\ket{l},\ket{r}$. 
 \fig{fig:JF} shows $J$ and $F$ (which  also has an analytic expression [SI]).  
  The inter-dot separation results in double-slit-like interference as phonons interact with the localised states  causing  oscillations  in $J$ and $F$, with a spectral period $\approx 2\pi/d^*$ \cite{stace:106801,Roulleau2011aa}.  This leads to a low frequency cut-off, $\sim1/d^*$, in $J$.  
Between the low- and high-frequency cutoff, $J$ is Ohmic with a superimposed oscillatory modulation.

\begin{figure}[t]
\begin{center}
\includegraphics[width=\columnwidth]{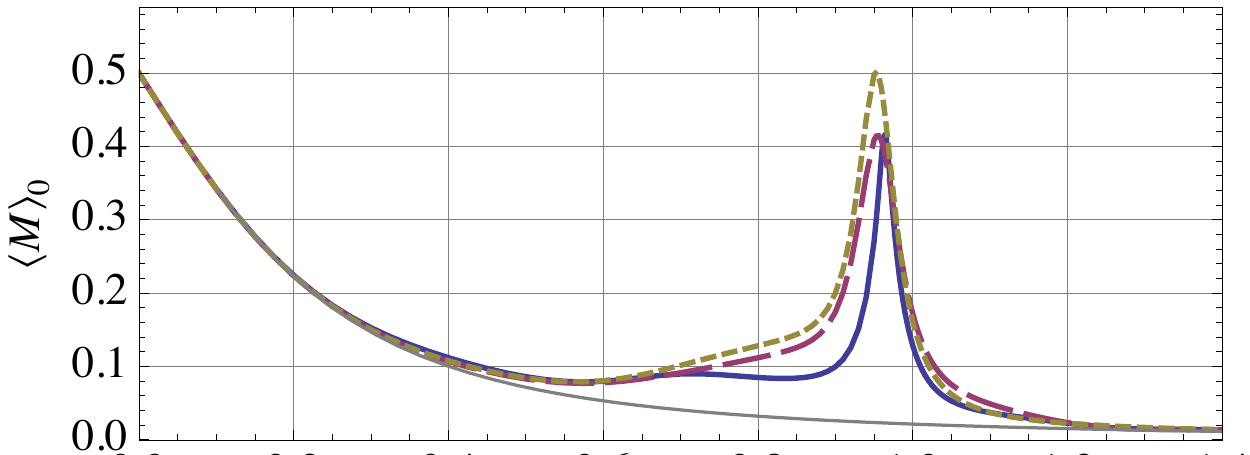}\vspace{-0.65mm}
\includegraphics[width=\columnwidth]{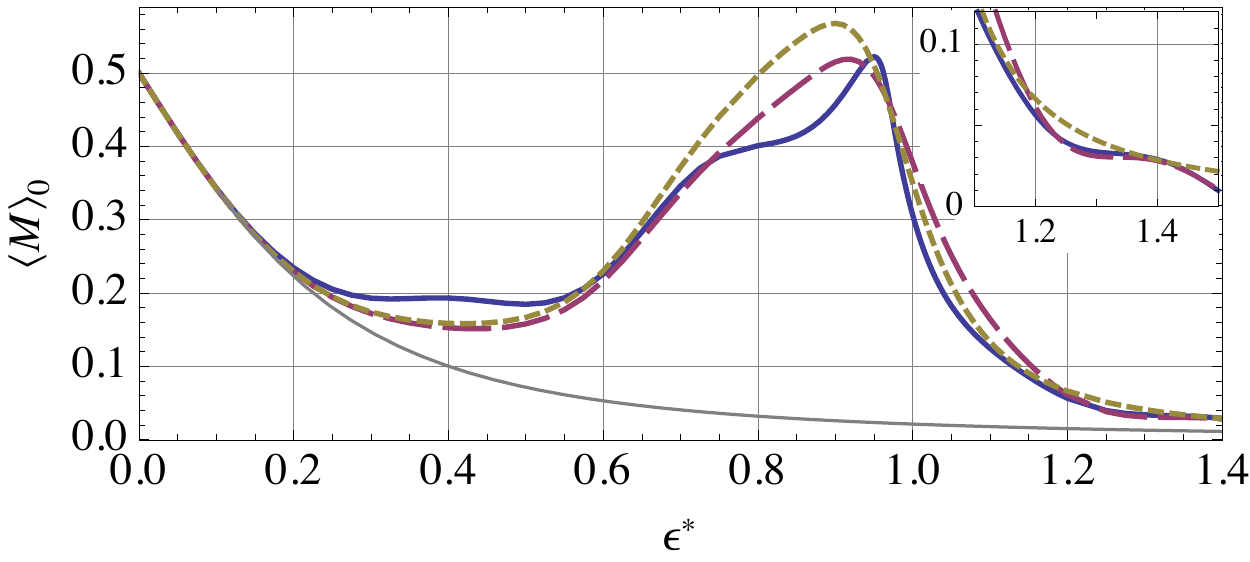}
\caption{Top: Time-average steady-state population of the right dot, $\langle M\rangle_0=\langle \ket{r}\bra{r}\rangle_0$, for relatively weak driving, $\Omega_0^*=0.07$.  Bottom: $\langle M\rangle_0$ for  stronger driving, $\Omega_0^*=0.2$. Inset: Zoom-in of red-detuned wing. Different curves correspond to different levels of approximation. Solid curve: $\mathcal{V}=\{0,\pm\Omega'\}$ and renormalised $\eta, \Omega$. Dashed curve: $\mathcal{V}=\{0,\pm\Omega'\}$ and bare $\tilde\eta, \Omega_{\textrm{approx}}$. Dotted curve: $\mathcal{V}=\{0\}$ and bare  $\tilde\eta, \Omega_{\textrm{approx}}$.    Light curve: No driving. Other parameters as in \fig{fig:etaomega}.} \label{fig:M}
\end{center}
\end{figure}

$F'(0)$  determines the renormalisation strength. 
We can calculate $F'(0)$ exactly [SI], but a good estimate is obtained by noting that  $F'(0)=-\pi^{-1}\int d\omega J(\omega)/\omega^2$.  Combined with the facts that $J\approx\pi\mathsf{P}|\omega|$ for $\omega\in[ -\omega_c^*,-1/d^*]$ and it decays rapidly outside this interval we find that $F'(0)\approx -\mathsf{P} \log(d^*\omega_c^*)<0$. 
Thus  the high- and low-frequency cut-offs appear as a ratio in the renormalisation strength.  It follows from \eqn{eqn:approxetaomega} that near resonance, the renormalised detuning is larger than the bare value, whilst the renormalised Rabi frequency is smaller.  

\fig{fig:etaomega} shows the effects of renormalisation on  $\eta$ and $\Omega$.  The renormalised quantities (solid curves) are modified most strongly close to resonance. 
 Near resonance, the renormalised detuning is steeper  than the bare detuning (dashed curve), consistent with the argument above. 
The renormalised Rabi frequency (bottom panel, solid curve) is lower than the bare value (dashed curve), and is well approximated by \eqn{eqn:approxetaomegaB} near resonance. 

The theoretical framework discussed here differs from  Markovian models in two respects: it includes contributions arising from (a) the dynamical  steady-state of the driven system, and (b) the bath-induced renormalisation of the system in which $H_D$ is chosen self-consistently to remove dispersive shifts and define the poles in $\mathcal{W}$.  We elucidate these contributions by manually suppressing each effect, and comparing with the fully dynamical, renormalised result.  We illustrate this by calculating the right dot population, $M=\ket{r}\bra{r}=(1-\sigma_z)/2$ \cite{barrett2006cmm}, which models an electrometer adjacent to the DQD \cite{petta:186802,reilly:162101}.    

\fig{fig:M} shows the time-averaged, steady-state population, $\langle M\rangle_0$, for relatively weak driving (top) and relatively strong driving (bottom).  The solid curves show the dynamical, renormalised results; the dashed curve retains the dynamical poles but neglects renormalisation \footnote{We use the rescaled Rabi frequency $\Omega_\textrm{approx}$ so that the effective driving amplitude is comparable near resonance.}; 
and the dotted curve neglects both, corresponding to the strong-driving Markov approximation \cite{stace:106801}. The resonant peaks exhibit strong phonon-induced asymmetry, which becomes more pronounced at higher driving, to the extent of exhibiting population inversion on the blue-detuned side \cite{Dykman79,stace:106801}.  The  enhancement of the blue-detuned wing is a consequence of photon absorption  from the driving field accompanied by a Raman phonon emission, leading to a higher  rate of excitation compared to the relaxation rate  \cite{stace:106801}.  At weak driving, the resonant peak is unsaturated ($\langle M\rangle_0<0.5$), \fig{fig:M} (top), but this is only evident when  dynamical poles are included (solid \& dashed curves); suppressing dynamical poles necessarily yields a saturated peak on resonance, i.e.\ $\langle M\rangle_0=0.5$ at $\eta=0$, (dotted curves).  
Consistent with \fig{fig:etaomega}, the effects of parameter renormalisation are most significant near resonance, narrowing the central resonant peak.  This occurs for two reasons: as $\epsilon$ moves away from resonance the renormalised detuning changes more rapidly with $\epsilon$ than does the bare detuning, and the renormalised Rabi frequency decreases below its resonant value.  

Following \eqn{eqn:residues2} we noted that there are residual $F$-dependent terms in \eqn{eqn:residues2} that cannot be cancelled.  This is manifest as  subtle `shoulders'  appearing on the red-detuned side of the resonance ($\phi>\omega_0$), as shown in the inset to \fig{fig:M}(bottom), resulting in non-Lorentzian decay of the red-detuned wings.  In this regime, the microwave photons have insufficient energy to drive real transitions between the energy eigenstates, highlighting the fact that the shoulders are a consequence of  dispersive, rather than dissipative,  electron-phonon coupling.

Phenomena such as the transition from  asymmetric unsaturated resonances at weak driving to population inversion  at strong driving, and  phonon-induced shoulders have been observed experimentally \cite{Colless2013aa}.  Our theory yields good qualitative agreement, and reasonable quantitative agreement with these experimental results.  Intriguing connections exist between our approach and a non-Markovian  extension to  Redfield theory 
\cite{JPSJ.74.3131,Ishizaki2008185}. 

In conclusion, we have derived a set of coupled equations for the residues of dynamical poles of the reduced density matrix of a system.  These terms  yield steady-state dynamics which are absent from Markovian treatments, as well as non-perturbative renormalisation of the bare system parameters.  Neglecting either the dynamical poles or the effects of renormalisation yields qualitatively different results, particularly near resonance. The theory is consistent with recent experimental results which exhibit  the same  bath-induced phenomena discussed here.  Our formalism  permits arbitrarily many Floquet eigenfrequencies in the driven Hamiltonian, so extends straightforwardly to  higher harmonics.  


This research was supported by the Office of the Director of National Intelligence, Intelligence Advanced Research Projects Activity (IARPA), through the Army Research Office grant W911NF-12-1-0354 and by the Australian Research Council via the Centre of Excellence in Engineered Quantum Systems (EQuS), project number CE110001013. 
We thank S.\ Barrett, J.\ Colless, J.\ Coombes, X.\ Croote, G.\ Milburn, A.\ Nazir and A.\ Shabani for illuminating discussions.

\bibliography{bib}

\end{document}